\documentclass[12pt]{book}

\usepackage[dvips]{graphicx,color}
\usepackage{makeidx,tsukuba}

\makeauthorindex
\makeindex

\def\Im{{\rm Im}\hskip2pt}
\def\simleq{\; \raise0.3ex\hbox{$<$\kern-0.75em \raise-1.1ex\hbox{$\sim$}}\; }
\def\simgeq{\; \raise0.3ex\hbox{$>$\kern-0.75em \raise-1.1ex\hbox{$\sim$}}\; }

\begin{document}

\BookTitle{\itshape The 28th International Cosmic Ray Conference}
\CopyRight{\copyright 2003 by Universal Academy Press, Inc.}
\pagenumbering{arabic}

\chapter{
Graviton Production by a Thermal Bath}

\author{Dario Grasso\\
{\it Scuola Normale Superiore, P.zza Dei Cavalieri 7, I-56126 PISA, Italy}
}

\section*{Abstract}
Thermal fluctuations in the early universe plasma and in very hot
astrophysical objects are an unavoidable source of gravitational waves (GW).
Differently from previous studies on the subject, we approach this
problem using methods based on field theory at finite temperature.
Such an approach allows to probe the infrared region of the spectrum
where dissipative effects are dominant. Incidentally, this region
is the most interesting from the point of view of the detectability
perspectives. We find significant deviations from a Planck spectrum.

\section{Introduction}

It is well known that thermal collisions in a plasma give rise to
production of gravitational radiation \cite{Weinberg}. Graviton emission 
is a consequence of particle acceleration in the scattering process 
(gravitational Bremsstrahlung). The emitted power of a non-relativistic
gas was determined by Weinberg \cite{Weinberg2} by 
summing up the radiated energies per collision under the condition
that the wave frequency is much larger than the mean collision frequency
($\omega \gg \omega_c$). This is required for the waves do not interfere.
As was noted in \cite{Weinberg}, for  $\omega \ll \omega_c$ the gas behaves
as a fluid rather than as a collection of independent particles.
The low frequency limit is the most interesting from the point of 
view of GW detection perspective. 
In this contribution we address the problem of computing the gravitational
emission rate by a relativistic plasma in the low frequency limit. 
We focus on the contribution to the emission process of collective
excitations of the plasma. For definiteness we consider here the case of a QED
plasma.

\section{Collective excitations in a relativistic plasma}

Fluctuations of electric and magnetic fields in a heat bath are related 
to the dispersive properties of the medium by the 
{\it fluctuation-dissipation theorem} \cite{Taji,Lemoine}. In a field theory language
this theorem states that the photon spectral function $\langle A_\mu A_\nu 
\rangle(q)$ is
related to the retarded and advanced Green's functions through
\begin{equation}
\langle A_\mu A_\nu \rangle(q) \equiv  i (1 - e^{\beta q_0})^{-1}
\left[ {\cal D}_{\mu\nu}(q_0+i\epsilon,{\bf q}) - 
{\cal D}^*_{\mu\nu}(q_0+i\epsilon,\bf{q})  \right]~,
\label{FDT}
\end{equation}
where $q$ is the photon 4-momentum and $\beta=1/T$ is the inverse temperature.
In the Feynman gauge this quantity can be decomposed into transversal and
longitudinal components
${\cal A}_{\mu\nu}=- P_{\mu\nu}{\cal A}_T- Q_{\mu\nu}{\cal A}_L$ 
where
\begin{equation}
    {\cal A}_{T,L}(q)=-\frac {1}{\pi}\,
        \frac{\Im \Pi_{T,L}}
        {|q^2-{\rm Re} \Pi_{T,L}|^2+|\Im \Pi_{T,L}|^2}~~
\label{ATL}.
\end{equation}
The expressions of  the transversal (longitudinal) components, $\Pi_{T(L)}$, 
of the polarization tensor, as well as the definitions 
of the projectors $P_{\mu\nu}$ and $Q_{\mu\nu}$, are given in 
\cite{Altherr}).
The dispersion relation of transverse (longitudinal) modes is given
by $q^2_o - {\bf q}^2 = {\rm Re} \Pi_{T(L)}(q)$. At small momenta 
$\Pi_{T(L)}(q)$ is of the order of the plasma frequency which for a 
relativistic plasma is $\omega_p \simeq eT/3$. The propagation of photons is 
strongly altered for $q_0,~{\bf q} \simleq \omega_p$; in this case they 
behave as quasi-particles endowed with an effective mass of order $eT$.
What is more relevant in the present context is that electromagnetic plasma 
fluctuations below  the light cone ($q_0 < |{\bf q}|$) are subject to
Landau damping. Graviton emission by electromagnetic excitations
take place because of the damping. The effect is analogous to axion production
due to the Primakoff effect (see e.g. \cite{Altherr}) and the results we present
here can be derived following an approach similar to that used to determine
the axion production rate in hot stars. 

Using Eqs.(\ref{FDT},\ref{ATL}) one finds \cite{Lemoine}
\begin{eqnarray}
\langle{\bf B}^2\rangle_q&=&\frac{2\pi}{1-e^{-\beta q_0}}
        2q^2{\cal A}_T(q)~~,\nonumber \\
\langle{\bf E}^2\rangle_q&=&\frac{2\pi}{1-e^{-\beta q_0}}
        \left[2q_0^2{\cal A}_T(q)+q^2{\cal A}_L(q)\right]~~.
\end{eqnarray}
In the following we will focus on magnetic fluctuations since electric fields
are screened in the static limit. 
Interestingly, it was showed by the authors of Refs.\cite{Lemoine,Taji} 
that a large peak is present in the magnetic fluctuations spectrum 
at nearly zero-frequency. This can be interpreted as ``squeezing''
of energy toward $q_0 = 0$  because of the non-linear coupling at
small momenta. The effect give rise to a random quasi-static magnetic field
with root-mean-square energy at the scale $l$ given by
\begin{equation}
B_l \simeq B_0 \left(\frac{l_p}{l}\right)^{3/2}~,\quad
B_0 = \left(\frac{T \omega_p^3}{32 \pi^{9/2}}\right)^{1/2}\; ,
\end{equation}
where $l_p \equiv 2\pi/\omega_p $.

\section{Gravitational waves production}

It is well known that stochastic magnetic fields can power gravitational 
waves production \cite{Durrer}. Magnetic anisotropic stresses 
\begin{equation}
\tau^{(B)}_{ij}({\bf k}) = \frac{1}{4\pi} \int d^3q~\left(B_i({\bf q})
B^*_j({\bf k-q}) - \frac {1}{2} B_l({\bf q}) B^{l~*}({\bf k-q})\delta_{ij} 
\right)
\end{equation}
act as a source term in the time evolution equation for the spatial
components of the metric perturbations.  In the Fourier space this 
equation can be written
\begin{equation}
\label{gweq}
{\ddot h}_{ij}  +  k^2 h_{ij} = 8\pi G \Pi^{(B)}_{ij}~,
\end{equation}
where
$\Pi^{(B)}_{ij} = \left(P_i^aP_j^b - \frac {1}{2} P_{ij}P_{ab}\right)~
\tau^{(B)}_{ab}$ 
is the tensor component of $\tau^{(B)}_{ij}$.
Due to the stochastic nature of thermal fluctuations, hence also of the GW
background they give rise to, the relevant quantity we are concerned here is
the GW spectral function $\langle h^{ij}({\bf k},t), h_{ij}({\bf k'},t) 
\rangle $.  Starting from the general solution of (\ref{gweq}) 
$\displaystyle h_{ij}({\bf k},t) = \frac {8\pi G}{k} \int_o^t dt'
\Theta(t - t') \sin[k(t - t')] \Pi_{ij}({\bf k'},t')$  and replacing 
ensemble average with time average, one finds \cite{noi,Koso}
\begin{equation}
\langle \: h_{ij} ({\bf k}, t) \: h_{ij} ^* ({\bf k} ', t)
\: \rangle \simeq
\frac{9 \sqrt{2} \: (16 \pi G)^2 \: \Delta t_*  }{16 \: k^3}
\delta^3 ({\bf k} - {\bf k} ') \int \! d^3 q \: B^2(q) B^2(|{\bf k} - {\bf q}|)
\end{equation}
where $\Delta t$ is the source duration.
Assuming a power-law spectrum $B^2(k) = A k^{n}$ for the fluctuations, 
it can be showed \cite{Caprini,Durrer} that whenever $n > -3/2$ the 
convolution is dominated by the frequency independent term 
$\displaystyle  \int \! d^3 q \: B^2(q) B^2(|{\bf k} - {\bf q}|) \simeq
4\pi A^2 \frac{k_{max}^{2n+3}}{2n+3}$. This give rise to a white-noise GW 
spectrum. In the previous section we showed that the energy density in
quasi-static magnetic fluctuations scales with the scale $l$ as $B^2_l \propto
l^{-3}$. Since $B^2_l \sim B^2(k) k^3|_{k=1/l} \propto l^{-(n+3)}$ it follows   
that $n = 0$ in this case, which fulfills the $n > -3/2$ condition. 
It was recently claimed by the authors of \cite{Caprini} that $n = 0$ is unphysical
and it should be replaced by $n = 2$. Such substitution would not affect our
results.  
In our case $k_{max}$ has to be identified with the plasma frequency $\omega_p$
and $k_{min}$ (see below) with the Hubble expansion rate $H(T)$.

The real-space correlator is given by 
\begin{equation}
\langle \: h_{ij} ({\bf x}, t) \: h^{ij} ({\bf x}, t)
\: \rangle =
\frac{V^2}{(2\pi)^6} \int \! d^3k d^3 k' e^{i({\bf k'}-{\bf k})\cdot {\bf x}}
\langle \: h_{ij} ({\bf k}, t) \: h^{ij~*}  ({\bf k} ', t)
\: \rangle  \; .
\end{equation}
The power-spectrum normalization requires 
$\displaystyle A = \frac{B^2_0}{8\pi}\frac{\pi^2}{V}
(n+3)~\omega_p^{-(n+3)}$. Then, for $n = 0$ we find
\begin{equation}
\langle \: h_{ij} ({\bf x}, t) \: h_{ij} ({\bf x}, t)
\: \rangle =  \frac{81 \sqrt{2}}{32} \: G^2 \: \Delta t  B_0^4 \omega_P^{-3}
\delta^3 ({\bf k} - {\bf k} ') \int_H^{\omega_P/2\pi} \!\frac{dk}{k} \;.
\end{equation}
The GW power spectrum is usually parametrized in terms of the  
{\em characteristic amplitude} $h_c(f)$ of GWs at frequency $f$,
defined by \cite{Maggiore} 
$\displaystyle \langle \: h_{ij} ({\bf x}, t) \: h_{ij} 
({\bf x}, t)\: \rangle \equiv 2 \int_0 ^\infty \frac{df}{f} h^2 _c (f,t)$.
The characteristic amplitude measured today at a frequency $f$ is
\begin{equation}
h_c (k_0,t_0) = \frac{a}{a_0} h_c ( k = \frac{a_0}{a} k_0 , t) 
\simeq  10^{-16}
        \left( \frac{100}{g_*} \right) ^{1/3}
        \left( \frac{\rm TeV}{T} \right)\; .
\end{equation}
It is natural to assume the emission time interval to be
$\Delta t \sim H^{-1} = g_*^{- 1/2} \frac{M_P}{ T ^2}$,
where $M_P$ is the Planck mass and $T$ the emission temperature. 
Disregarding the weak dependence on the number of relativistic d.o.f. $g_*$
we finally find
\begin{equation}
\label{result}
h_c (f,t_0) \simeq 10^{-24} \kappa^3 \left (\frac{T}{M_P}\right)^3 \; ,
\end{equation} 
where we parametrized the plasma frequency dependence on $T$ as $\omega_P =
\kappa T$. For QED $\kappa \simeq 0.1$ but it can be larger for GUT theories.
Furthermore the larger number of gauge bosons predicted by these theories
may very well give rise to more than a factor 10 enhancement in the predicted 
value of $h_c$.
The minimal wavenumber today corresponds to the red-shifted value of $H(T)$.
In term of frequency it is $\displaystyle f_{min} \simeq 10^{-4}~{\rm Hz} 
\left( \frac{T}{{\rm TeV}} \right)$. 
It is remarkable that the expected GW signal may be in the detectable range of 
LISA \cite{LISA} if gravity becomes strong ($M_P \sim TeV$) at the TeV scale
(A specific model should be adopted here. If, for example, modification of
gravity take please because of the presence of large extra-dimensions, 
the role of gravitons escaping in the bulk should be taken into account).  

Independently on these speculations, our main result is to have shown that
collective excitations in a relativistic plasma give rise to a GW background
with a frequency independent spectrum. At low frequencies this background 
overwhelm that produced by the conventional       
gravitational bremsstrahlung.
\section*{Acknowledgments}
\noindent
The author would like to thank D. Comelli, R. Durrer, T. Kahniashvili,
 A. Nicolis and M. Pietroni for valuable discussions.

\endofpaper
\end{document}